\documentclass[12pt]{article}

\usepackage{amsfonts,amsmath,amssymb}
\usepackage{hyperref}
\usepackage{cite}
\usepackage{epsfig}
\usepackage{paralist}
\usepackage{fancyhdr}
\usepackage{tikz}
\usepackage{tkz-euclide}
\usetikzlibrary{decorations.pathmorphing,patterns,calc,snakes,arrows}

\usepackage{graphicx}
\usepackage{xcolor}
\numberwithin{equation}{section}
\usepackage[vcentermath]{youngtab}
\usepackage{etex}
\usepackage{braket}
\usepackage{float}

\def\spa#1{\phantom{\fbox{\rule[-#1cm]{0cm}{0cm}}}}

\def\be{\begin{equation}}
\def\ee{\end{equation}}
\def\bea{\begin{eqnarray}}
\def\eea{\end{eqnarray}}

\renewcommand{\thefootnote}{\fnsymbol{footnote}}

\makeatletter
\g@addto@macro\bfseries{\boldmath}
\makeatother

\def\Tb{{\bar{T}}}

\begin{document}

\hfuzz=100pt
\title{{\Large \bf{Nonperturbative effects in $T\Tb$-deformed conformal field theories: A toy model for Planckian physics}}}
\date{}
\author{Shinji Hirano$^{a, b}$\footnote{
	e-mail:
	\href{mailto:shinji.hirano@gmail.com}{shinji.hirano@gmail.com}}
  \,and Vinayak Raj$^{a}$\footnote{
	e-mail:
	\href{mailto:vinayak.hep.th@gmail.com}{vinayak.hep.th@gmail.com}}}
\date{}

\maketitle

\thispagestyle{fancy}
\rhead{YITP-25-113}
\cfoot{}
\renewcommand{\headrulewidth}{0.0pt}

\vspace*{-1cm}
\begin{center}
$^{a}${{\it School of Science, Huzhou University}}
\\ {{\it Huzhou 313000, Zhejiang, China}}
  \spa{0.5} \\
$^b${{\it Center for Gravitational Physics and Quantum Information (CGPQI)}}
\\ {{\it  Yukawa Institute for Theoretical Physics, Kyoto University}}
\\ {{\it Kitashirakawa-Oiwakecho, Sakyo-ku, Kyoto 606-8502, Japan}}
\spa{0.5}  

\end{center}

\begin{abstract}
We propose a nonperturbative completion of two-point correlators in $T\Tb$-deformed conformal field theories (CFTs), and analyze their behavior at distance scales shorter than the fundamental length scale set by the $T\Tb$ deformation. Building on the interpretation of the $T\Tb$ deformation as a coupling to two-dimensional quantum gravity with a unique built-in length scale, we advance the study of $T\Tb$-deformed CFTs as a toy model for Planckian physics.
As we probe shorter distances, trans-Planckian oscillations are followed by a super-Planckian regime in which correlations are typically suppressed by geometric randomness, in contrast to the power-law growth characteristic of CFTs. Moreover, their dependence on distance becomes exponentially weaker, suggesting that the underlying geometric structure has been largely erased --  a behavior broadly consistent with expectations for quantum spacetime in this regime.
\end{abstract}

\renewcommand{\thefootnote}{\arabic{footnote}}
\setcounter{footnote}{0}

\newpage

\tableofcontents


\section{Introduction}
\label{Sec:Introduction}

The challenge of formulating General Relativity (GR) as a quantum field theory (QFT) stems from the fact that in four spacetime dimensions, Newton's constant $G_N$ has the dimension of length squared. This dimensionality implies that gravity is an irrelevant interaction in the renormalization group (RG) sense: its coupling grows with energy. Without some mechanism to tame this growth -- either via asymptotic safety \cite{Weinberg:1978,Weinberg:1980gg} or through a UV completion that introduces new degrees of freedom, such as strings -- GR generically leads to uncontrollable divergences at short distances.

Although such proposals remain under active investigation, it is instructive to explore simplified or lower-dimensional models in which gravitational features can be made explicit and tractable. In this spirit, the $T\bar{T}$ deformation of two-dimensional quantum field theories \cite{Zamolodchikov:2004ce, Smirnov:2016lqw, Cavaglia:2016oda} provides a unique setting: it induces a coupling to two-dimensional gravity with a built-in length scale, rendering the geometry effectively dynamical~\cite{Dubovsky:2017cnj, Dubovsky:2018bmo,Tolley:2019nmm}. This allows us to study physical observables such as  correlators in a controlled, solvable framework that exhibits some features reminiscent of quantum gravity, including the modification of short-distance behavior and the breakdown of local QFT structure at high energies.

Despite the solvability of $T\bar{T}$-deformed theories in the energy spectrum and S-matrix~\cite{Smirnov:2016lqw,Cavaglia:2016oda}, a nonperturbative understanding of correlation functions -- especially at distances shorter than the deformation scale --  has remained elusive even in light of the notable progress made in~\cite{Aharony:2023dod, Barel:2024dgv}  and~\cite{Cui:2023jrb,Chen:2025jzb}.\footnote{Important progress has been made in understanding the nonperturbative structure of $T\bar{T}$-deformed partition functions.
A possible form of the nonperturbative completion of the $T\bar{T}$-deformed CFT partition function was first explored in~\cite{Aharony:2018bad}.
In parallel, exact nonperturbative results were obtained in the context of $T\bar{T}$-deformed gauge theories in earlier works~\cite{Griguolo:2022xcj,Griguolo:2022hek}.
Subsequently, the nonperturbative structure of the torus partition function was further developed in~\cite{Gu:2024ogh,Gu:2025tpy}. } 
This gap is especially significant because short-distance observables are sensitive to the breakdown of locality and may encode signatures of quantum gravitational behavior.

In this work, we propose a nonperturbative completion of two-point correlators in $T\Tb$-deformed CFTs and analyze their behavior across all distance scales. We find a crossover from a trans-Planckian regime, characterized by oscillatory behavior, to a super-Planckian phase in which correlations are suppressed by geometric randomness. Moreover, their dependence on distance becomes exponentially weaker, suggesting that the underlying geometric structure has been largely erased -- a behavior broadly consistent with expectations for quantum spacetime in this regime.

These results build on the interpretation of the $T\bar{T}$ deformation as coupling to two-dimensional quantum gravity, where the presence of a fundamental length scale softens UV behavior through dynamical geometry. The framework provides a rare, solvable setting in which features of Planckian physics -- such as nonlocality, geometric disorder, and correlation suppression -- emerge from a controlled quantum field theoretic deformation.

The remainder of this paper is organized as follows. In Section \ref{sec:QG}, we review the interpretation of the $T\bar{T}$-deformed conformal field theory (CFT) as a two-dimensional theory of quantum gravity, emphasizing the presence of a fundamental length scale and the role of dynamical geometry. In Section \ref{sec:NP}, we construct a nonperturbative completion of two-point correlators by resumming the leading logarithmic contributions, yielding exact expressions valid across all distance scales. Section \ref{sec:Planckian} analyzes the behavior of these correlators in the short-distance regime, revealing a transition from trans-Planckian oscillations to a regime where correlations are suppressed and their dependence on distance becomes exponentially weaker -- a hallmark of coarse-grained or effectively nonlocal spacetime structure. In Section \ref{sec:EE}, we examine the entanglement entropy derived from the two-point functions of twist operators, uncovering nonperturbative effects and a potential signature of a minimal length scale. Finally, Section \ref{sec:discussion} concludes with a discussion of the implications of our results for Planckian physics and possible directions for future research.

\section{$T\bar{T}$-deformed CFT as a theory of quantum gravity}
\label{sec:QG}

We advance the view that the $T\bar{T}$-deformed CFT admits an interpretation as a two-dimensional quantum gravity theory characterized by a unique length scale $\ell_P=\sqrt{|\mu|}$. To this end, we adopt Tolley's massive gravity formulation~\cite{Tolley:2019nmm}, defined by the action
\begin{align} \label{FT_action}
S_{T\bar{T}}[e, f]
= S_{\rm CFT}[e] + \frac{1}{2\mu} \int d^2x \epsilon^{ij} \epsilon_{ab} \left(e^a_i - f^a_i\right)\left(e^b_j - f^b_j\right)
\equiv S_{\rm CFT}[e] + S_{\rm mG}[e, f]\ .
\end{align}
Here, $e$ and $f$ are zweibeins corresponding to the two-dimensional metrics
$ds_{\rm CFT}^2 = h_{ij} dx^i dx^j$ and
$ds_{T\bar{T}}^2 = \gamma_{ij} dx^i dx^j$, respectively.
Crucially, the metric $h_{ij}$ (or equivalently, the zweibein $e^a_i$) is dynamical, thereby promoting the theory to a version of two-dimensional quantum gravity. The second metric $\gamma_{ij}$ describes a fixed background geometry on which the $T\bar{T}$-deformed theory is defined.

The observables of the $T\bar{T}$-deformed CFT are related to those of the undeformed CFT through the generating functional of the deformed theory, given by
\begin{align}
Z_{T\bar{T}}[f, J] = \int \mathcal{D}e^a_i e^{-S_{\rm mG}[e, f]} Z_{\rm CFT}[e, J]\ ,
\end{align}
where $J$ collectively denotes the sources for operators in the theory. This functional satisfies a diffusion-type or ``Schr\"odinger'' equation,
\begin{align} \label{qFlow}
\frac{\partial}{\partial \mu} Z_{T\bar{T}}[f, J]
= \int d^2x \lim_{y \to x} \frac{1}{2} \epsilon_{ij} \epsilon^{ab}
\frac{\delta^2 Z_{T\bar{T}}[f, J]}{\delta f^a_i(y) \delta f^b_j(x)}\ ,
\end{align}
which arises from the random geometry interpretation of the $T\bar{T}$ deformation at infinitesimal coupling~\cite{Cardy:2018sdv}.

This brief review offers a conceptual foothold, emphasizing that we are dealing with a theory of quantum gravity and, in what follows, will explore the effects of fluctuating geometry around the Planck scale $\ell_P = \sqrt{|\mu|}$.

\section{Nonperturbative completion of two-point correlators}
\label{sec:NP}

To quantitatively assess the impact of geometric fluctuations, we focus on the two-point correlators of the operator ${\cal O}_{\Delta}$ with conformal dimension $\Delta$ in a $T\bar{T}$-deformed CFT. Restricting our attention to the leading logarithmic corrections, we begin with the expression derived in~\cite{HR_Correlators_mGrav}, obtained by integrating out the dynamical zweibein $e^a_i$ within the massive gravity formulation of $T\bar{T}$-deformed theories:
\begin{equation}
\begin{aligned} \label{2pt_all_orders_mG}
\hspace{-.4cm}
\langle {\cal O}_{\Delta}(x_1){\cal O}_{\Delta}(x_2)\rangle^{\text{leading-log}}_{T\bar{T}}
&= \frac{\Gamma(-\Delta+1)}{\pi 2^{2\Delta}\Gamma(\Delta)} \int_{-\infty}^{+\infty} d^2k |k|^{2(\Delta-1)} \exp^{i\vec{k} \cdot \vec{x}_{12} + \frac{\mu}{\pi}|k|^2 \ln(|x_{12}|/\varepsilon)}\ ,
\end{aligned}
\end{equation}
where $\varepsilon$ is a reference scale, which we later identify with the Planck scale $\ell_P$ of this quantum gravity theory. This expression holds for generic values of $\Delta$, 
excluding $\Delta\in \mathbb{Z}_+$, which will be treated separately. 
Expanding this expression in powers of the $T\bar{T}$ coupling $\mu$ reproduces the known all-order perturbative result~\cite{Cardy:2019qao} (see also~\cite{Hirano:2020nwq, Hirano:2024eab}):
\begin{equation}
\begin{aligned} \label{2pt_all_orders_pert}
\hspace{-.4cm}
\langle {\cal O}_{\Delta}(x_1){\cal O}_{\Delta}(x_2)\rangle^{\text{leading-log}}_{T\bar{T}}
&= \sum_{n=0}^{\infty} (-1)^n \frac{(4\mu)^n}{n! \pi^{n}} \prod_{k=0}^{n-1} (\Delta + k)^2
\frac{\ln^n(|x_{12}|/\varepsilon)}{|x_{12}|^{2(\Delta+n)}}\ .
\end{aligned}
\end{equation}
First, the perturbative series, viewed as a function of 
\begin{align}
Z \equiv -\frac{4\mu}{\pi |x_{12}|^2} \ln(|x_{12}|/\varepsilon),
\end{align}
is non-convergent and at best formal. To make sense of it -- particularly in the short-distance regime $|Z| \gg 1$ -- a nonperturbative completion is required.

As it turns out, when $Z < 0$, the perturbative series~\eqref{2pt_all_orders_pert} is Borel summable. For $Z > 0$, although the Borel transform develops a singularity, the integrals in~\eqref{2pt_all_orders_mG} can still be evaluated, yielding a closed-form, nonperturbative expression. (For related background, see, for example, the pedagogical review on resurgence theory~\cite{Marino_resurgence}.) We now proceed to elaborate on the details of these two cases.

To set our notation, we denote the perturbative series in~\eqref{2pt_all_orders_pert} by $f(Z)$. Its Borel transform takes the form of a Gauss hypergeometric function:
\begin{align}
{\cal B}[f](\xi)={1\over |x_{12}|^{2\Delta}}\sum_{n=0}^{\infty} {(\Delta)_n(\Delta)_n\over (1)_nn!} \xi^n
={1\over |x_{12}|^{2\Delta}}\mbox{}_2F_1(\Delta, \Delta; 1; \xi)
\end{align}
where $(\Delta)_n=\Gamma(\Delta+n)/\Gamma(\Delta)$ is the Pochhammer symbol. The Borel transform exhibits a singularity at $\xi = 1$. However, for $Z < 0$, the integration variable satisfies $\xi =Zt\le 0$, and the perturbative series for the two-point correlator becomes Borel summable, yielding
\begin{equation}
\begin{aligned}\label{NP_2pt_Zn}
{\cal S}[f]=\int_0^{\infty}e^{-t}{\cal B}[f](Zt)
&={(-Z)^{-\Delta}\over |x_{12}|^{2\Delta}}U\left(\Delta, 1, -1/Z\right)\qquad\mbox{for}\qquad Z<0\ ,
\end{aligned}
\end{equation}
where $U(a,b,z)$ is the Tricomi confluent hypergeometric function of the second kind, which admits the integral representation
\begin{align}
U(a,b,z)={1\over\Gamma(a)}\int_0^{\infty}dt e^{-zt}t^{a-1}(1+t)^{b-a-1}\ .
\end{align}
It is straightforward to verify that ${\cal S}[f]$ admits the asymptotic expansion~\eqref{2pt_all_orders_pert} for small $Z$, corresponding to the long-distance regime.
Note also that since ${\cal S}[f]$ in~\eqref{NP_2pt_Zn} is well-defined for $\Delta \in \mathbb{Z}_+$, this nonperturbative expression remains valid for all values of $\Delta \ge 0$.

We now turn to the case $Z > 0$. The singularity on the Borel plane signals the presence of an instanton-like sector. At first glance, it is not obvious how to access this nonperturbative contribution. Fortunately, however, the integral in~\eqref{2pt_all_orders_mG} can be evaluated directly.
We begin by considering generic values of $\Delta$. 
By shifting and rescaling momentum variables as $\vec{q}=|x_{12}|(\vec{k}-\vec{a})$ with $\vec{a}=-i\pi|x_{12}|\vec{x}_{12}/(2\mu\ln(|x_{12}|/\varepsilon))$, the expression~\eqref{2pt_all_orders_mG} can be rewritten as 
\begin{equation}
\begin{aligned}\label{2pt_mu_negative_SD}
\langle {\cal O}_{\Delta}(x_1){\cal O}_{\Delta}(x_2)\rangle^{\rm leading{\text -}log}_{T\Tb}
&={ \Gamma\left(-\Delta+1\right)\over \pi 2^{2\Delta}\Gamma\left(\Delta\right)|x_{12}|^{2\Delta}}e^{-{1\over Z}}
\int_{-\infty}^{+\infty}d^2q\left|\vec{q}+\vec{a}\right|^{2(\Delta-1)}e^{-{Z\over 4}|q|^2}\ .
\end{aligned}
\end{equation}
Using the identity
\begin{align}
\left|\vec{q}+\vec{a}\right|^{2(\Delta-1)}={1\over\Gamma(1-\Delta)}\int_0^{\infty}dt\, t^{-\Delta}
e^{-t\left|\vec{q}+\vec{a}\right|^2}\ ,
\end{align}
and performing the Gaussian integral over $\vec{q}$, followed by a change of integration variable, we obtain
\begin{equation}
\begin{aligned}
\langle {\cal O}_{\Delta}(x_1){\cal O}_{\Delta}(x_2)\rangle^{\rm leading{\text -}log}_{T\Tb}
&={Z^{-\Delta}e^{-{1\over Z}}\over \Gamma\left(\Delta\right)|x_{12}|^{2\Delta}}
\int_0^{\infty}du\,  {u^{-\Delta}\over u+1}\, e^{u\over Z(u+1)}\ .
\end{aligned}
\end{equation}
Introducing the variable $s = u/(u+1)$, one recognizes the integral as a representation of the Kummer confluent hypergeometric function, yielding\footnote{It is noteworthy that both hypergeometric functions appearing in the nonperturbative completion satisfy the second-order differential equation
\begin{align}
x F'' +(1-x)F' -(1-\Delta) F=0
\end{align}
whose general solution can be written as
\begin{align}
F(x)=\alpha\, \mbox{}_1F_1(1-\Delta; 1; x)+\beta \left(U(1-\Delta, 1, x)+e^xU(\Delta, 1, -x)\right)\ ,
\end{align}
for arbitrary constants $\alpha$ and $\beta$, where the two hypergeometric functions are related by the identity
\begin{align}
U(1-\Delta, 1, x)=e^x\Gamma(\Delta)\biggl[{U(\Delta, 1, -x)\over\Gamma(1-\Delta)}+(-1)^{\Delta}\mbox{}_1F_1(\Delta; 1; -x)\biggr]\ .
\end{align}
}
\begin{equation}
\begin{aligned}\label{negative_mu_NP_complete_2pt}
\langle {\cal O}_{\Delta}(x_1){\cal O}_{\Delta}(x_2)\rangle^{\rm leading{\text -}log}_{T\Tb}
&={\Gamma(1-\Delta) \over |x_{12}|^{2\Delta}}e^{-{1\over Z}}Z^{-\Delta}\mbox{}_1F_1(1-\Delta; 1; 1/Z)\quad\mbox{for}\quad Z>0\ .
\end{aligned}
\end{equation}
Strictly speaking, the integral representation is convergent only in the range $0<{\rm Re}\Delta<1$. However, since the hypergeometric function itself is well-defined for generic  $\Delta$, we analytically continue the result from this range to arbitrary values of  $\Delta$, excluding $\Delta\in\mathbb{Z}_+$.
For small $Z$, this expression admits a trans-series expansion:
\begin{equation}
\begin{aligned}\label{negative_mu_perturbative}
\hspace{-.5cm}
\langle {\cal O}_{\Delta}(x_1){\cal O}_{\Delta}(x_2)\rangle^{\rm leading{\text -}log}_{T\Tb}
&={1\over |x_{12}|^{2\Delta}}\sum_{n=0}^{\infty}{(\Delta)_n(\Delta)_n\over n!}Z^n\\
&+{(-1)^{\Delta-1}\Gamma(1-\Delta) \over \Gamma(\Delta)|x_{12}|^{2\Delta}}e^{-{1\over Z}}Z^{1-2\Delta}\sum_{n=0}^{\infty}{(1-\Delta)_n(1-\Delta)_n\over n!}(-Z)^n\ ,
\end{aligned}
\end{equation}
where the first line reproduces the perturbative series~\eqref{2pt_all_orders_pert}, while the second line captures the advertised instanton-like sector associated with the Borel singularity, which is exponentially suppressed in the long-distance regime $Z\ll 1$.

We next turn to the special case of $\Delta\in \mathbb{Z}_+$. In this case, the momentum-integral representation~\eqref{2pt_all_orders_mG} is modified to
\begin{equation}
\begin{aligned}
\hspace{-.3cm}
\langle {\cal O}_{\Delta}(x_1){\cal O}_{\Delta}(x_2)\rangle^{\rm leading{\text -}log}_{T\Tb}
={(-1)^{\Delta}\over\pi 2^{2\Delta-1}\Gamma(\Delta)^2}\int_{-\infty}^{+\infty}d^2k|k|^{2(\Delta-1)}\ln|k|e^{i\vec{k}\cdot\vec{x}_{12}+{\mu\over\pi}|k|^2\ln(|x_{12}|/\varepsilon)}.
\end{aligned}
\end{equation}
We omit the intermediate steps, as they involve only a minor variation from the generic $\Delta$ case. We find
\begin{equation}
\begin{aligned}\label{negative_mu_NP_complete_2pt_integer}
\hspace{-.0cm}
\langle {\cal O}_{\Delta}(x_1){\cal O}_{\Delta}(x_2)\rangle^{\rm leading{\text -}log}_{T\Tb}
&={(-1)^{\Delta}e^{-{1\over Z}}\over 4^{\Delta}\Gamma\left(\Delta\right)^2}
\partial_{\Delta}\biggl[{\Gamma(\Delta)\over |x_{12}|^{2\Delta}}\left({4\over Z}\right)^{\Delta}\mbox{}_1F_1\left(1-\Delta; 1; {1\over Z}\right)\biggr]
\end{aligned}
\end{equation}  
for the special values of $\Delta\in \mathbb{Z}_+$.

To complete the construction of the nonperturbative two-point correlators, we must match the two expressions valid in different regimes at the point $Z=0$.
There are two values of $|x_{12}|$ for which $Z$ vanishes: at large distances, $|x_{12}|/\sqrt{|\mu|}\to \infty$, and at the Planck scale $|x_{12}|=\ell_P$, where we identify $\varepsilon=\ell_P$.
The matching point between the two expressions is the Planck scale. Since both take the form 
\begin{align}
\langle {\cal O}_{\Delta}(x_1){\cal O}_{\Delta}(x_2)\rangle^{\text{leading-log}}_{T\Tb}
\simeq {1\over |x_{12}|^{2\Delta}}\left(1+4\Delta^2Z+\cdots\right)
\end{align}
in the vicinity of  $Z=0$, this guarantees a smooth connection at the Planck scale. To summarize, we have established that 
\begin{align}\label{NP2pt_corrected}
\hspace{-.4cm}
\langle {\cal O}_{\Delta}(x_1){\cal O}_{\Delta}(x_2)\rangle^{\text{leading-log}}_{T\Tb}
&=
\begin{cases}
\dfrac{(-Z)^{-\Delta}}{|x_{12}|^{2\Delta}}  U\left(\Delta, 1, -\dfrac{1}{Z}\right) & \text{for } Z < 0 \\[1ex]
\dfrac{\Gamma(1-\Delta) Z^{-\Delta}}{|x_{12}|^{2\Delta}} e^{-{1\over Z}}  \mbox{}_1F_1\left(1-\Delta; 1; \dfrac{1}{Z}\right) & \text{for } Z > 0
\end{cases}
\end{align}
for generic (non-integer) values of $\Delta$, where the second line is replaced by~\eqref{negative_mu_NP_complete_2pt_integer} when $\Delta\in \mathbb{Z}_+$.

\section{Short-distance behavior and Planckian regimes}
\label{sec:Planckian}

We now investigate the short-distance physics of this toy model of quantum gravity, characterized by the fundamental length scale $\ell_P=\sqrt{|\mu|}$, using the nonperturbative two-point correlator in~\eqref{NP2pt_corrected} as a probe. For clarity, we reiterate the identification of the reference scale with the Planck scale:
\begin{align}
\varepsilon = \ell_P\equiv \sqrt{|\mu|}\ .
\end{align}
Under this identification, the variable $Z$ takes the form
\begin{align}
Z \equiv -{\rm sgn}(\mu)\frac{4\ell_P^2}{\pi |x_{12}|^2} \ln(|x_{12}|/\ell_P)
\end{align}
which parametrizes the relevant distance scales. 

The behavior of the two-point correlator across all distance scales can be summarized as follows:
At long distances, it asymptotically approaches the undeformed CFT result.
As one probes shorter distances, trans-Planckian oscillations set in, followed by a super-Planckian regime where correlations are typically suppressed by geometric randomness -- marking a stark departure from the power-law behavior characteristic of CFTs.\footnote{Short-distance suppression of correlators was previously noted in~\cite{Aharony:2023dod, Barel:2024dgv, Cui:2023jrb, Chen:2025jzb}. However, while we focus on standard local operators defined on the ``worldsheet,'' their analyses concern local operators in the ``target space,'' which generally appear non-local from the worldsheet perspective. As a result, their main conclusions regarding short-distance behavior differ from ours, although certain qualitative features are shared.}
In this regime, the correlator's dependence on distance becomes exponentially weaker, indicating that the underlying geometric structure has been largely erased -- a phenomenon broadly consistent with expectations for quantum spacetime near the Planck scale.

\begin{figure}[h!]
\centering \includegraphics[height=2.4in]{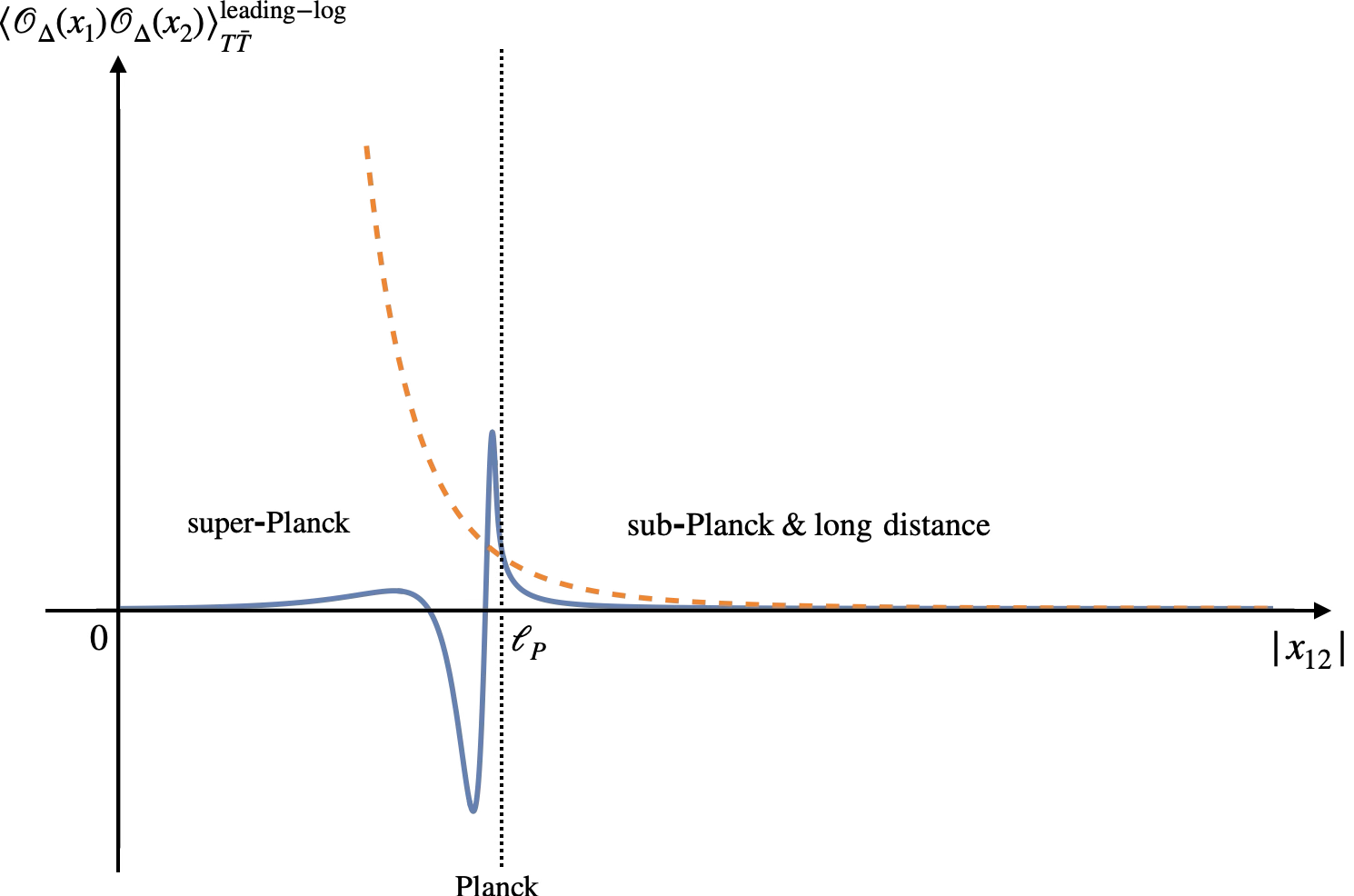}
\caption{Behavior of the two-point correlator (with $\Delta=5/2$) for $\mu>0$ corresponding to the case in  which the energy becomes complex at short distances or for large conformal dimensions. The correlator exhibits oscillations at trans-Planckian scales and is suppressed by inverse powers of logarithms at super-Planckian scales.  As we increase $\Delta$, the oscillations become more rapid. For comparison, the dashed orange curve shows the undeformed CFT two-point correlator.}
\label{fig:2pt_plot_cplx_E}
\end{figure}  
\begin{figure}[h!]
\centering \includegraphics[height=2.4in]{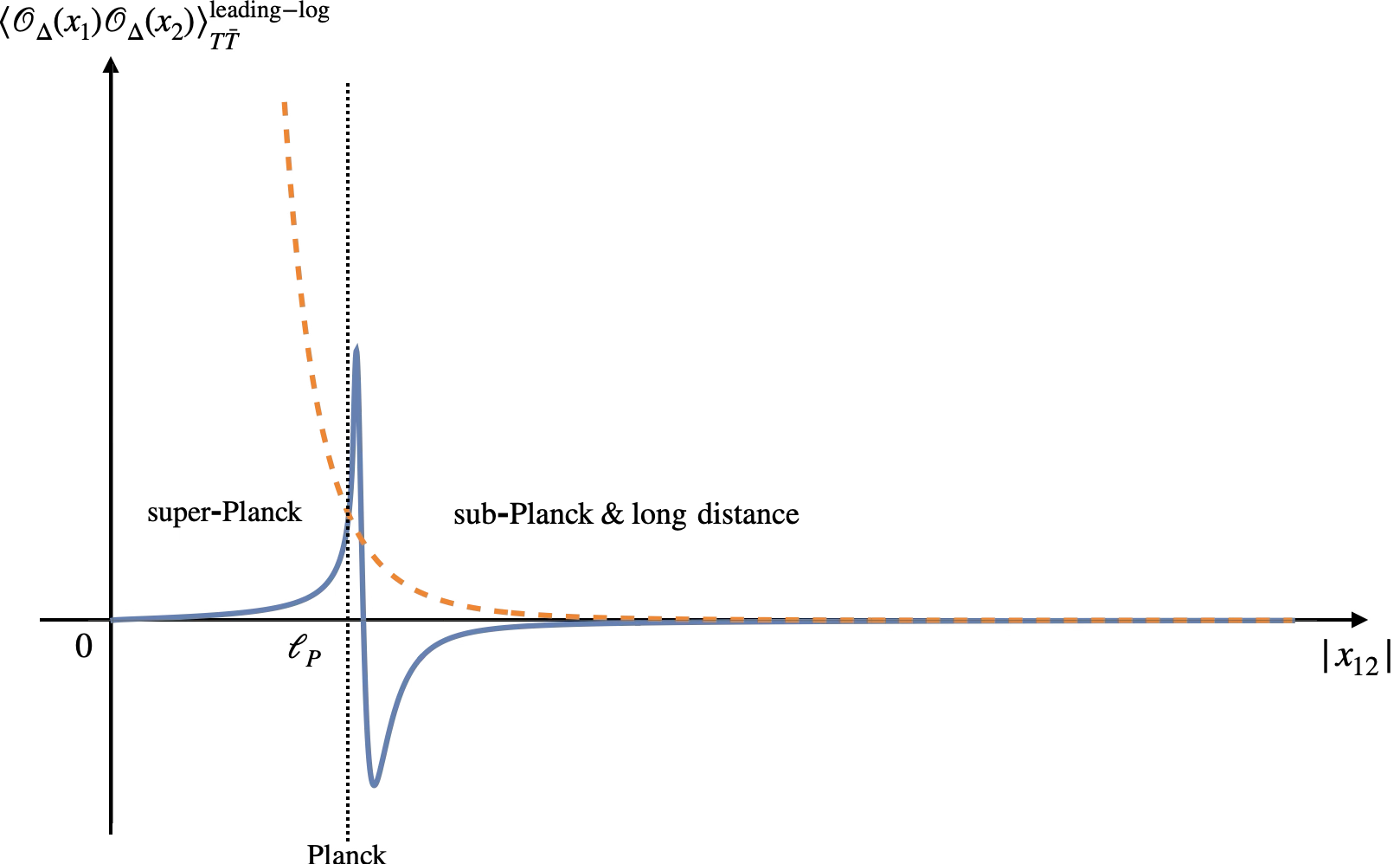}
\caption{Behavior of the two-point correlator with $\Delta = 5/2$ for $\mu < 0$. In this case, the theory exhibits a Hagedorn phase at high temperatures, and states with low conformal dimensions become unstable at short distances. The correlator displays oscillatory behavior near the Planck scale in the sub-Planckian regime and is suppressed by a power of logarithm at ultra-short distances. For comparison, the dashed orange curve shows the undeformed CFT two-point correlator.}
\label{fig:2pt_plot_Hagedorn}
\end{figure}  
\begin{figure}[h!]
\centering \includegraphics[height=2.4in]{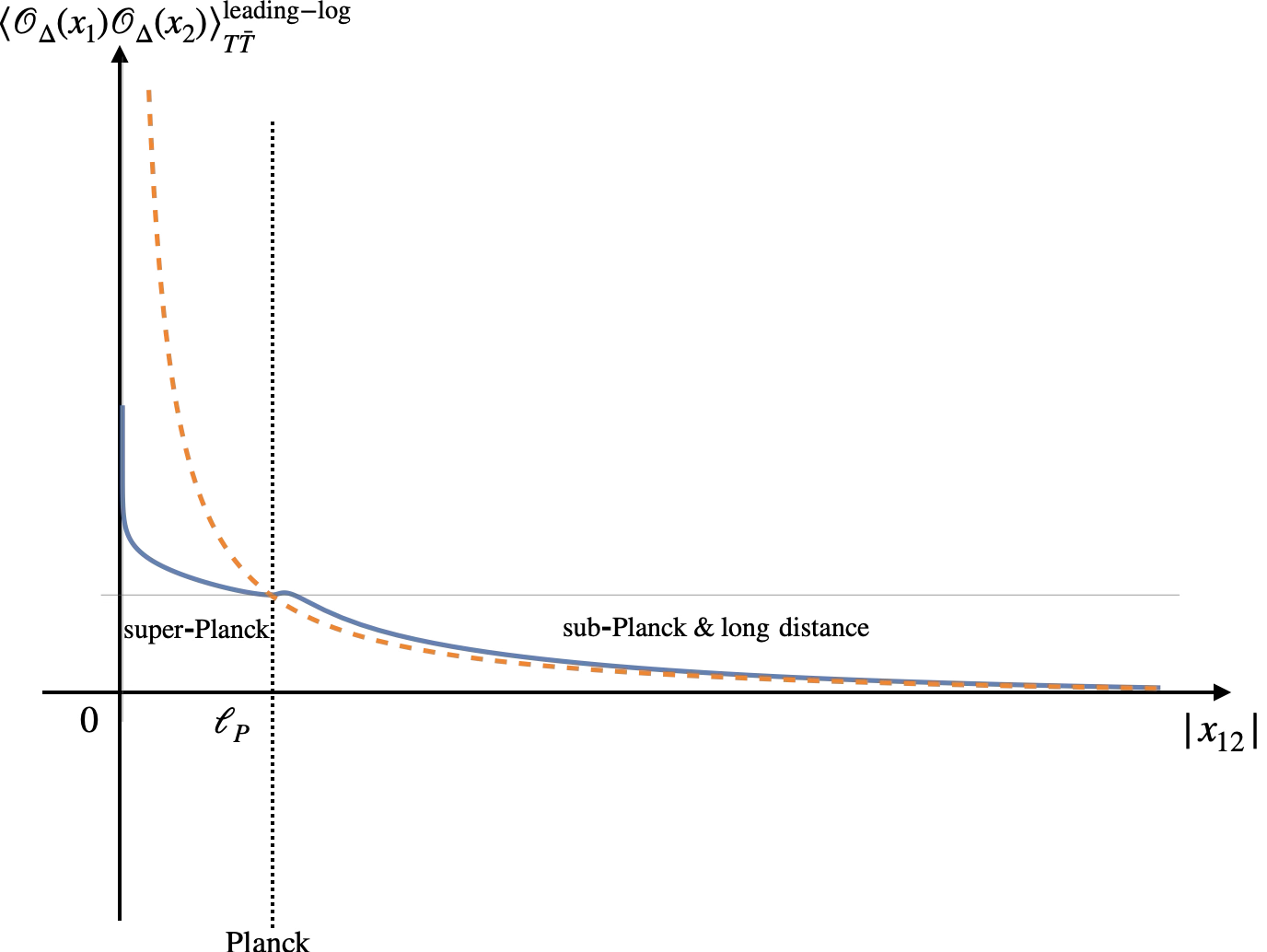}
\caption{Behavior of the two-point correlator with low conformal dimension ($\Delta = 1/2 < 1$) for $\mu < 0$. The correlator diverges logarithmically as $|x_{12}|/\ell_P \to 0$. However, the divergence remains exponentially milder than that of the undeformed CFT correlator. For reference, the dashed orange curve shows the CFT two-point correlator.}
\label{fig:2pt_plot_Hagedorn_low}
\end{figure}  
Typical correlation patterns across all distance scales are shown in Figures \ref{fig:2pt_plot_cplx_E}--\ref{fig:2pt_plot_Hagedorn_low}, categorized by the sign of the $T\bar{T}$ coupling $\mu$ and the conformal dimension $\Delta$. More precisely, super-Planckian suppression is observed in all cases except when $\Delta < 1$ with $\mu < 0$. Even in these exceptional cases, however, the growth of correlations remains exponentially milder than that of the undeformed CFT.
To make this behavior more explicit, we present the analytic expressions for the two-point correlators in the super-Planckian regime, where $|x_{12}| \ll \ell_P$:
\begin{align}\label{NP2pt_corrected_short_distance}
\langle {\cal O}_{\Delta}(x_1){\cal O}_{\Delta}(x_2)\rangle^{\text{leading-log}}_{T\Tb}
&\sim\begin{cases}
\dfrac{\Gamma(1 - \Delta)}{\left(4\ell_P^2 \ln(\ell_P/|x_{12}|)\right)^{\Delta}} 
& \text{for }\quad \mu > 0 \vspace{0.5em}\\
\dfrac{\left(\pi/4\ell_P^2\right)^{\Delta} }{\Gamma(\Delta)\left(\ln(\ell_P/|x_{12}|)\right)^{\Delta - 1}} 
& \text{for }\quad \mu < 0 \ .
\end{cases}\end{align}
These results hold for generic (non-integer) values of $\Delta$ when $\mu > 0$, and for all values of $\Delta$ when $\mu < 0$.
In the special case $\Delta \in \mathbb{Z}_+$ with $\mu > 0$, the correlator takes the form
\begin{equation}
\begin{aligned}
\langle {\cal O}_{\Delta}(x_1){\cal O}_{\Delta}(x_2)\rangle^{\text{leading-log}}_{T\Tb}
&\sim {(-)^{\Delta-1}\pi^{\Delta}\ln\ln(\ell_P/|x_{12}|)\over \Gamma\left(\Delta\right)\left(4\ell_P^2\ln(\ell_P/|x_{12}|)\right)^{\Delta}}\ .
\end{aligned}
\end{equation}
Beyond the suppression, as advertised, a particularly striking feature is that the correlator's dependence on distance becomes exponentially weaker.
For generic $\Delta$ with positive $\mu$, this behavior suggests that the effective notion of distance is nearly constant, modified only by a logarithmic correction:\footnote{A similar interpretation applies to other cases as well, with minor variations in the form of the logarithmic correction.}
\begin{align}\label{distance_change}
|x|^2\quad\longrightarrow\quad \ell_P^2\ln(\ell_P/|x|)\ .
\end{align}
In other words, the underlying geometric structure has been largely erased. Were it entirely washed out, the effective distance would reduce to a constant. As such, we believe this behavior represents the closest possible realization of the expectations for quantum spacetime in this regime.

\section{Entanglement entropy}
\label{sec:EE}

It is of interest to investigate how quantum entanglement responds to fluctuations in the underlying geometry, particularly when the entangling surfaces themselves are subject to fluctuation.\footnote{For a discussion of the $T\bar{T}$ deformation of entanglement entropy on a cylinder, see~\cite{Chen:2018eqk}. Meanwhile, the $T\bar{T}$ deformation of R\'enyi entropy has been analyzed perturbatively to first order in~\cite{He:2019vzf, He:2020qcs}. More recently,~\cite{Lai:2025thy} investigates nonperturbative aspects of R\'enyi entropy in $T\bar{T}$-deformed CFTs, highlighting a potentially important role for instanton-like contributions.}
Given the nonperturbative control of two-point functions, this question naturally presents itself as the next step to explore. As we will see, however, the results are somewhat puzzling: in one case, they broadly align with physical intuition, while in the other, they exhibit behavior that diverges markedly from what one might have anticipated.
There is, however, an important caveat: our analysis is restricted to the leading logarithmic contribution, and subleading corrections might alter the overall picture.

\begin{figure}[h!]
\centering \includegraphics[height=3.0in]{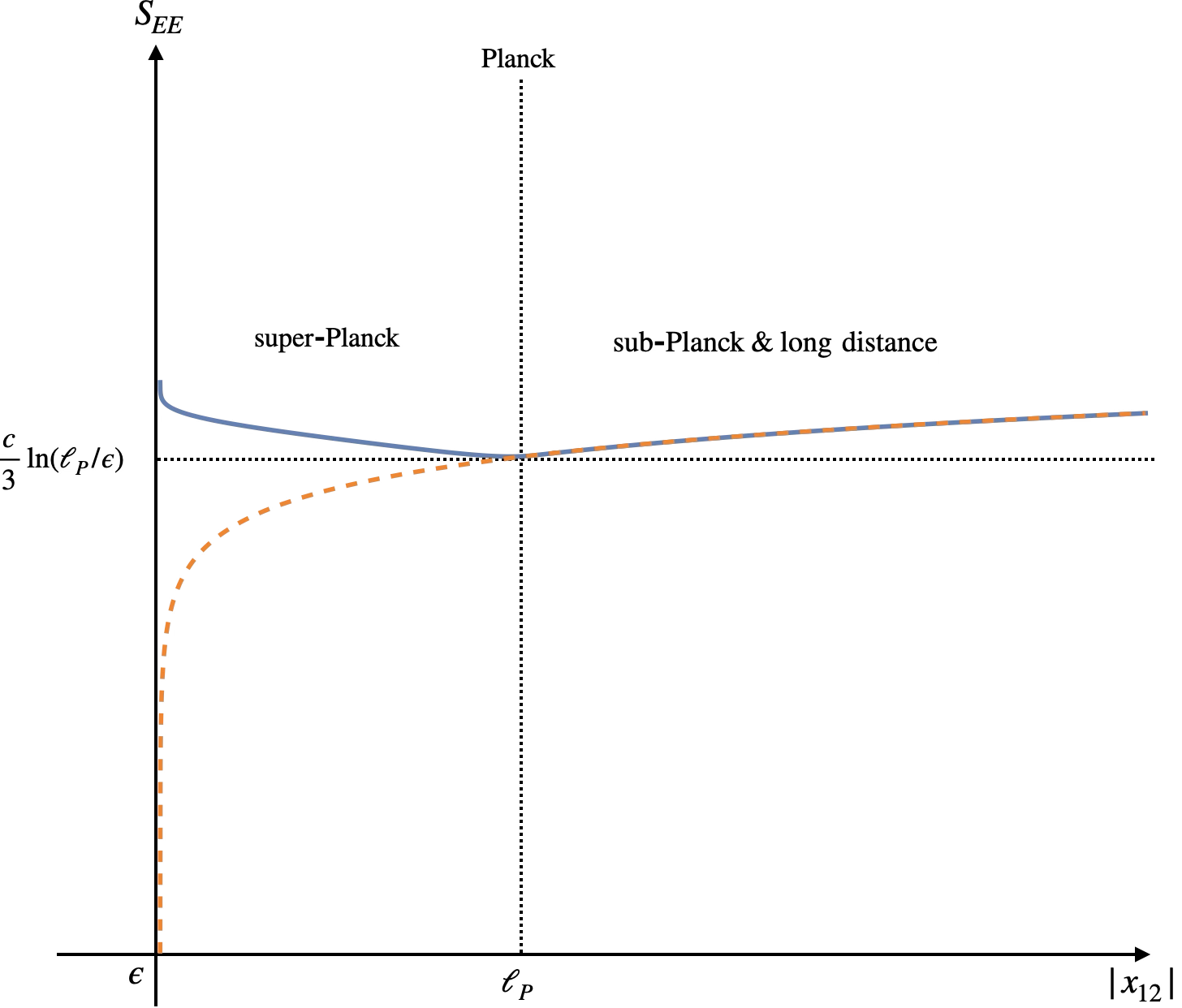}
\caption{Entanglement entropy (blue curve) for the interval between $x_1$ and $x_2$ in the case $\mu > 0$. The dashed orange curve shows the entanglement entropy of the undeformed CFT for comparison. Notably, the entanglement entropy reaches a minimum at the Planck scale and begins to increase mildly as the interval becomes shorter, reflecting the modified notion of effective distance described in~\eqref{distance_change}. This suggests that the Planck scale may represent a minimal length in a physical sense. The observed behavior arises entirely from the instanton-like sector.}
\label{fig:EE_cplx_E}
\end{figure}

\begin{figure}[h!]
\centering \includegraphics[height=3.0in]{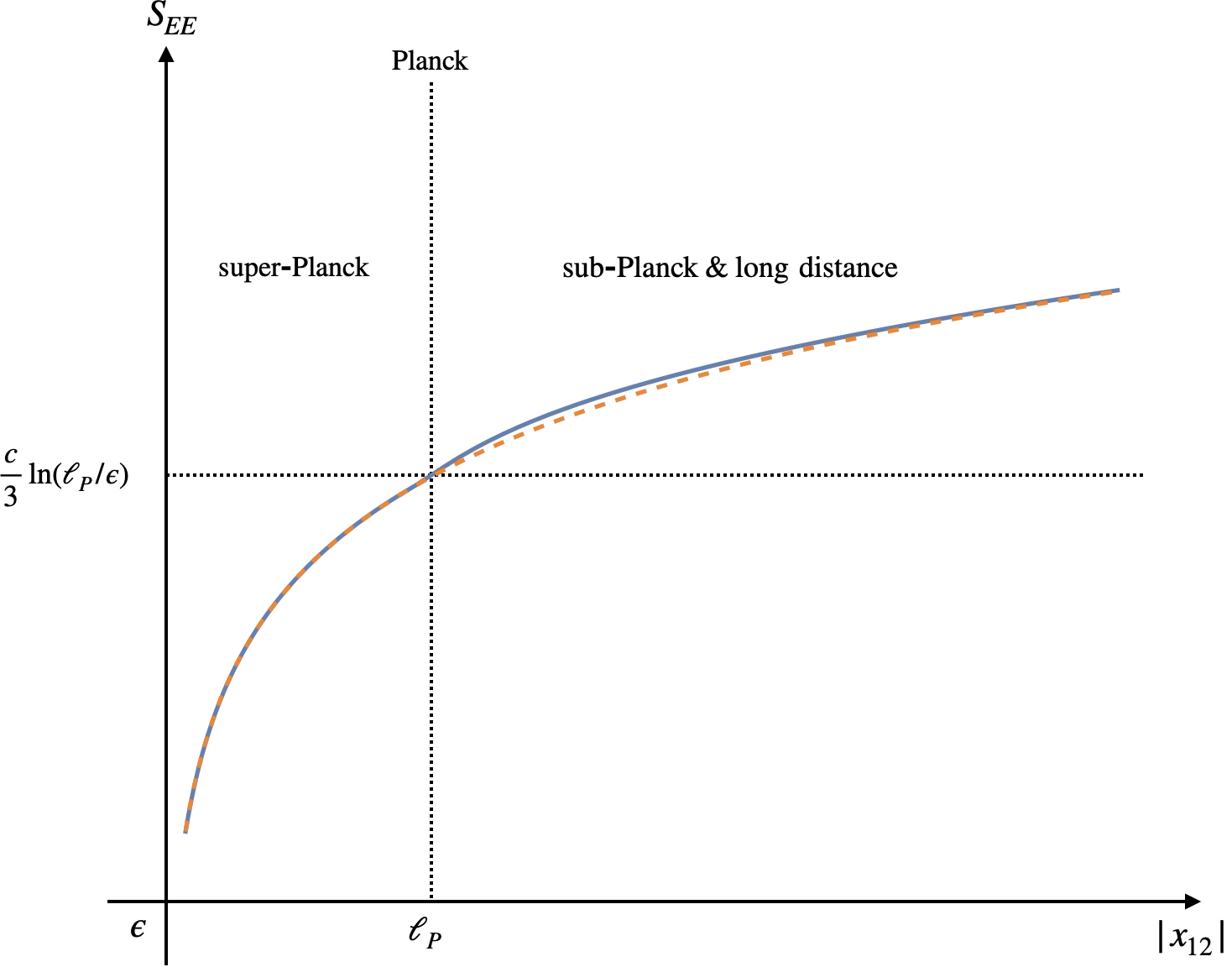}
\caption{Entanglement entropy (blue curve) for the interval between $x_1$ and $x_2$ in the case $\mu < 0$. The dashed orange curve shows the corresponding result in the undeformed CFT. In this case, the behavior of the entanglement entropy exhibits no significant qualitative deviation from that of the CFT.}
\label{fig:EE_Hagedorn}
\end{figure}

The entanglement entropy can be computed by evaluating the two-point correlator of twist operators with conformal dimension  $\Delta={c\over 12}\left(n-{1\over n}\right)$:
\begin{align}
S_{EE}=-{c\over 6}\lim_{\Delta\to 0}\partial_{\Delta}\left(\epsilon^{2\Delta}\langle {\cal O}_{\Delta}(x_1){\cal O}_{\Delta}(x_2)\rangle^{\text{leading-log}}_{T\Tb}\right)\ ,
\end{align}
where $\epsilon$ is a UV regulator (not to be confused with $\varepsilon$).
Introducing the notation
\begin{align}
\langle {\cal O}_{\Delta}(x_1){\cal O}_{\Delta}(x_2)\rangle^{\text{leading-log}}_{T\Tb}={1\over |x_{12}|^{2\Delta}}F(\Delta; Z)\ ,
\end{align}
the entanglement entropy becomes
\begin{align}
S_{EE}={c\over 3}\ln(|x_{12}|/\epsilon)-{c\over 6}\lim_{\Delta\to 0}\partial_{\Delta}F(\Delta; Z)\ .
\end{align}
It is worth emphasizing that the $T\bar{T}$ corrections to the perturbative two-point correlator (at leading-log order) are proportional to $\Delta^2$ at all orders. Consequently, the entanglement entropy receives no contribution from the perturbative sector; any $T\bar{T}$ correction to the EE originates purely from nonperturbative effects.
As discussed in the previous section, there are two nonperturbative forms of the function $F(\Delta; Z)$:
\begin{align}
F(\Delta;Z)=\begin{cases}
(-Z)^{-\Delta}U\left(\Delta, 1, -1/Z\right) & \mbox{for}\quad Z<0 \\
\Gamma(1-\Delta) e^{-{1\over Z}}Z^{-\Delta}\mbox{}_1F_1(1-\Delta; 1; 1/Z) & \mbox{for}\quad Z>0
\end{cases}\ .
\end{align}
In the case of positive $\mu$, the first form applies in the sub-Planckian regime, while the second governs the behavior at super-Planckian scales. For negative $\mu$, the roles of the two forms are reversed.
In the first case, the $T\bar{T}$ corrections vanish, as seen from
\begin{align}
\lim_{\Delta\to 0}\partial_{\Delta}\left((-Z)^{-\Delta}U\left(\Delta, 1, -1/Z\right)\right)=-\ln(-Z)-\ln(-1/Z)=0\ ,
\end{align}
which reflects the absence of any $T\bar{T}$ contribution from the perturbative sector. By contrast, in the second case, the presence of the instanton-like sector gives rise to a nonvanishing correction to the entanglement entropy. The resulting entanglement entropy is shown for both signs of $\mu$ in Figures~\ref{fig:EE_cplx_E} and~\ref{fig:EE_Hagedorn}.

For $\mu > 0$, one can derive an explicit analytic expression for the entanglement entropy in the super-Planckian regime $|x_{12}| \ll \ell_P$:
\begin{align}
S_{EE}\simeq {c\over 6}\ln\left[\left({\ell_P\over\epsilon}\right)^2\ln{\ell_P\over |x_{12}|}\right]-{c\over 6}(\gamma-\ln(4/\pi))\ ,
\end{align}
where the interval must remain larger than the UV regulator: $|x_{12}| \ge \epsilon$. 
The first term reflects the modified notion of effective distance described in~\eqref{distance_change}. More qualitatively, when the interval size $|x_{12}|$ falls below the Planck length $\ell_P$, fluctuations of the entangling surface (i.e., the interval endpoints) can cause the effective size to increase, leading to an enhancement of the entanglement entropy. 

In contrast, for $\mu < 0$ -- the case associated with a Hagedorn phase at high temperatures -- the entanglement entropy receives no $T\bar{T}$ correction below the Planck scale ($|x_{12}| < \ell_P$), as the instanton-like sector does not contribute in this regime. Nevertheless, a mild deviation from the CFT behavior is observed in the sub-Planckian region due to trans-Planckian oscillations. By contrast, the behavior in the super-Planckian regime -- seemingly unaffected by geometric fluctuations -- remains particularly puzzling. We leave the understanding of this feature to future investigation and debate.

\section{Discussion}\label{sec:discussion}

In this work, we have constructed a nonperturbative completion of two-point correlators in $T\bar{T}$-deformed CFTs and analyzed their behavior across all distance scales. At long distances, the correlators reduce to the familiar power-law form of the undeformed CFT. As we probe shorter distances, trans-Planckian oscillations set in, followed by a super-Planckian regime in which correlations are suppressed by geometric randomness and the dependence on distance becomes exponentially weaker. This qualitative transition signals the breakdown of locality and the emergence of an effectively coarse-grained geometry. Altogether, the results provide a concrete realization of several hallmark features expected of quantum spacetime near the Planck scale -- including a softened UV structure and a minimal length scale.

It has been speculated that quantum spacetime at super-Planckian scales behaves effectively as if it were two-dimensional~\cite{Atick:1988si, Ambjorn:2005db, Lauscher:2005qz, Modesto:2008jz, Horava:2009if} (see~\cite{Carlip:2009kf} for a comprehensive review of this idea). In light of this, our results may reflect universal features of quantum spacetime that extend beyond the two-dimensional setting. The emergence of a minimal length scale and the exponential weakening of correlations at short distances -- governed by an effectively coarse-grained geometry -- align with expectations from dimensional reduction scenarios in higher-dimensional quantum gravity. The $T\bar{T}$ deformation thus provides a tractable framework in which such phenomena are realized explicitly, suggesting that the behaviors we observe are not mere artifacts of two-dimensionality but may instead offer a concrete glimpse into the deep ultraviolet structure of quantum geometry.

Finally, the Hagedorn phase~\cite{Atick:1988si} remains one of the less understood regimes in high-energy physics. With the analytical control afforded by $T\bar{T}$-deformed theories, extending our analysis to finite temperatures may offer a concrete and tractable framework to explore the Hagedorn phase in detail, potentially shedding new light on its physical implications.


\section*{Acknowledgments}

SH would like to thank the department of mathematics at Nagoya University for their hospitalities during his visits where part of this work was done. The work of SH is supported in part by the National Natural Science Foundation of China under Grant No.12147219.


\end{document}